\documentclass[checkin,showpacs,aps,pra,onecolumn]{revtex4-1}

\usepackage{mathrsfs}

\usepackage{color,soul}

\newcommand{\bn}{\begin{enumerate}}
\newcommand{\en}{\end{enumerate}}
\newcommand{\ba}{\begin{eqnarray}}
\newcommand{\ea}{\end{eqnarray}}

\usepackage{graphicx,color}

\newcommand{\rr}{{\bf r}}

\newcommand{\be}{\begin{equation}}
\newcommand{\ee}{\end{equation}}

\newcommand{\et}{{\it et al. }}

\def\prl{{ Phys. Rev. Lett. }}

\def\pra{{ Phys. Rev. A }}

\begin{document}

\newcommand{\clr}{\color{black}}










\title{Magic high-order harmonics from a
  quasi-one-dimensional hexagonal solid}


\author{G. P. Zhang$^*$}

 \affiliation{Department of Physics, Indiana State University,
   Terre Haute, IN 47809, USA }

\author{Y. H. Bai}

\affiliation{Office of Information Technology, Indiana State
  University, Terre Haute, IN 47809, USA }

\date{\today}

\begin{abstract}
  { High-order harmonic generation (HHG) from atoms is a coherent
    light source that opens up attosecond physics, but it is the
    application of HHG to solids that brings much of excitement for
    the last decade. Here we report a completely new kind of harmonics
    in a quasi-one-dimensional and hexagonal barium titanium sulfide:
    Under circularly polarized laser excitation, harmonics are
    generated only at first, fifth, seventh and eleventh orders. These
    magic harmonics appear only with circularly polarized light, not
    with linearly polarized light. Neither cubic nor tetragonal cells
    have magic harmonics even with circularly polarized light.
    Through a careful group-theory analysis, we find that two
    subgroups of symmetry operations unique to the hexagonal symmetry
    cancel out third and ninth harmonics.  This feature presents a
    rare opportunity to develop HHG into a crystal-structure
    characterization tool for phase transitions between hexagonal and
    nonhexagonal structures.  }
\end{abstract}




 \maketitle

\section{Introduction}



\newcommand{\ik}{i{\bf k}}
\newcommand{\jk}{j{\bf k}}

\newcommand{\Ba}{BaTiS$_3$ }
\newcommand{\Bae}{BaTiS$_3$}

High-order harmonic generation (HHG) from solids \cite{farkas1992} and
nanostructures \cite{pronin1994,prl05,ganeev2009a,ganeev2013} has been
extended to a broad scope of materials (see references in
\cite{ghimire2011,nc18,kruchinin2018}). With strong signals and large
tunability of energy spectra, solid-state HHG has reshaped the
landscape of HHG as a radiation source from simple atoms, and has
gradually developed into a practical tool to characterize materials
properties on unprecedented short time scales, where the motion of
electrons can be pictured frame by frame within several hundred
attoseconds. Naturally, not all the processes such as chemical
reactions require such short time scales, but many do. For instance,
to resolve laser-induced ultrafast spin dynamics \cite{eric}, a short
pulse is necessary, since it allows one to disentangle the magnetic
and electronic dynamics from nuclear vibrational dynamics. However,
condensed matters are far more complex than atoms. The advantage of
HHG over other tools has not been materialized, though using HHG to
map bands has been proposed \cite{vampa2015a}. 

Recently, \Ba shows a broadband birefringence in infrared regions
\cite{niu2018}, but it is its quasi-one dimensionality and strong
optical anisotropy that caught our attention. Low-dimensional
materials with large oscillator strength are indispensable to
nonlinear optical responses, if their band gaps ($E_g$) are small but
nonzero. According to the well-known scaling rule, the third-order
susceptibility $\chi^{(3)}$ is proportional to $\chi^{(3)}\propto
E_g^{-6}$ \cite{agrawal1978,prb99}.  \Ba has a tiny gap, so even
within a perturbation limit, its nonlinear susceptibility is expected
to be strong, but little is known about its nonlinear optical
properties, and even less its high-order harmonic generation.

In this paper, we predict strange high-order harmonics in hexagonal
barium titanium sulfide (\Bae) that circularly ($\sigma$) polarized
light generates harmonics only at a few special orders.  Regardless of
laser pulse duration and photon energy, $\sigma$ light only induces
first, fifth, seventh and eleventh harmonics, not third and ninth
harmonics.  We call them magic harmonics.  Linearly ($\pi$) polarized
light only generates normal odd-order harmonics. This finding is
independent of whether the system has an inversion symmetry or
not. Neither cubic nor tetragonal systems have magic harmonic orders
even excited with $\sigma$ light.  We carry out a detailed group
symmetry analysis and find that magic harmonics are associated with
the hexagonal group symmetry
\cite{hellwarth1977,butcher,shang1987,boyd}.  This group contains two
subgroups: subgroup A contains the identity matrix and 180$^\circ$
rotation, and subgroup B includes the four proper rotations, $C_6$,
$C_6^2$, $C_6^4$, and $C_6^5$. Each subgroup only generates a normal
harmonic spectrum, but if they both are present, they generate a
destructive interference and exactly cancel out harmonics at the third
and ninth orders.  The same conclusion is found for the six improper
rotations. Since cubic or tetragonal systems do not have such magic
harmonics, magic harmonics found here present an opportunity to
develop HHG into a possible structure characterization tool for phase
transitions between hexagonal and nonhexagonal structures
\cite{izyumov} in varieties of materials
\cite{iwata2005,menniger1996,milledge1959,olander2003,halo2011,gomez-castano2018,saha2018,concas2011,ariz1997,castedo2011}. Our
finding complements the prior studies using linearly polarized light
\cite{ghimire2011,you2017,lakhotia2018} well.

The rest of the paper is arranged as follows. In Sec. II, we outline
our theoretical formalism. Our main results are presented in Sec. III,
where we provide the details of our structural optimization,
information of the electronic states, and high harmonic generations,
followed by a symmetry group analysis. Finally, we conclude this paper
in Sec. IV.

\section{Theoretical formalism}

\Ba is a quasi-one dimensional material with Ti and S atoms forming a
chain along the $c$ axis, with chain-chain distance of 6.749$\rm\AA$.
Figure \ref{fig1}(a) shows its crystal structure.  Ti chains pass
through face-sharing sulfur octahedra.  Figure \ref{fig1}(b) shows its
structure projected on to the $ab$ plane, where S atoms form a
distinctive hexagonal motif and Ba atoms fill the empty space left
behind. According to Huster \cite{huster1980} and Singh \et
\cite{niu2018}, \Ba adopts a hexagonal BaNiO$_3$ structure with space
group No. 194, $\rm P6_3/mmc$. This structure has an inversion
symmetry, with Ti at the Wyckoff position (2a), Ba at (2d) and S at
(6h). Table \ref{huster} shows the Huster's structure information.
However, Niu \et \cite{niu2018} suggested a different space group
No. 186, $\rm P6_3mc$, which has a lower symmetry without inversion
symmetry and the number of symmetry operations is reduced from 24 to
12. Which structure, Niu's or Huster's, is more stable is an open
question.

Theoretically, we employ the state-of-the-art density functional
theory \cite{wien2k} to optimize the \Ba structure, with little input
from the experiments.  We first solve the Kohn-Sham equation
\cite{wien2k,np09,prb09}, \be \left
     [-\frac{\hbar^2\nabla^2}{2m_e}+V_{ne}+V_{ee}+V_{xc} \right
     ]\psi_{\ik}(\rr)=E_{\ik} \psi_{\ik} (\rr), \label{ks} \ee where
     $m_e$ is the electron mass, the terms on the left-hand side
     represent the kinetic energy, nuclear-electron attraction,
     electron-electron Coulomb repulsion and exchange correlation
     \cite{pbe}, respectively.  $\psi_{\ik}(\rr)$ is the Bloch
     wavefunction of band $i$ at crystal momentum ${\bf k}$, and
     $E_{\ik}$ is the band energy.  We include the spin-orbit coupling
     (SOC) using a second-variational method in the same
     self-consistent iteration \cite{wien2k}, though we find the
     effect of SOC is very small.  Wien2k \cite{wien2k} employs the
     linearlized augmented planewave basis. In our calculation, the
     dimensionless product of planewave cutoff $K_{\rm max}$ and
     Muffin-tin radius $R$ is $RK_{\rm max}=9$.  Such a large value
     ensures that even higher eigenstates are accurately described.
     The Muffin-tin radius for each element is as follows, $R_{\rm
       mt}(\rm Ba)=2.5 $ Bohr, $R_{\rm mt}(\rm Ti)=2.32 $ Bohr, and
     $R_{\rm mt}(\rm S)=2.06 $ Bohr, so the core charges are confined
     within the spheres.  We use a {\bf k} mesh of $23\times 23\times
     24$, which is more than enough to converge our results.

To simulate HHG, we employ a laser pulse with duration 48 fs and
photon energy 1.6 eV. These laser parameters are commonly used in
experiments.  We numerically solve the time-dependent Liouville
equation for density matrices $\rho_{\bf k}$ at each ${\bf
  k}$ \cite{np09} \be i\hbar \frac{\partial \rho_{\bf k}}{\partial
  t}=[H,\rho_{\bf k}], \ee where $H$ contains both the system
Hamiltonian and the interaction between the laser and system.  The
expectation value of the momentum operator \cite{prl05,pra06} is
computed from \be {\bf P}(t)=\sum_{\bf k} {\rm Tr}[\rho_{\bf k}(t)
  \hat{{\bf P}}_{\bf k}],\ee where the trace is over band indices and
crystal momentum {\bf k}.  We include all the states from band 41 to
146 (see the arrows in Fig. \ref{fig4}(a)), which cover the major
portion of the energy spectrum.  Calculations using different parts of
the energy spectrum are also carried out, but there is no qualitative
difference.  To compute the harmonic signal, we Fourier transform
${\bf P}(t)$ to frequency domain (see details in Ref. \cite{nc18}),
\be {\bf P}(\Omega)=\int_{-\infty}^{\infty} {\bf P}(t) {\rm
  e}^{i\Omega t} {\cal W}(t) dt, \ee where ${\cal W}(t)$ is the window
function. Each component of ${\bf P}(\Omega)$ requires a separate
Fourier transform.  We find that the window function is necessary
since small oscillations in {\bf P}(t) at the end of the time window
easily hide the harmonic structures at high orders. We emphasize that
this window function does not alter the amplitude of the harmonic
signal.  We choose a hyper Gaussian ${\cal W}(t)=\exp[-b(at)^8]$,
where $a$ and $b$ determine the width of the window function and the
starting and ending times. In our current study, we use $a= 0.035$/fs
and $b = 5 \times 10^{-9}$ (no unit), which spans the entire region of
our data.

\section{Results}

\subsection{Structural optimization}

Since earlier studies by Huster \cite{huster1980}, structurally Niu
and coworkers \cite{niu2018} provided two possible structures for \Ba
with the same group symmetry.  According to the international tables
for crystallography \cite{hahn}, in group 186 ($\rm P6_3mc$), Ti
Wyckoff positions are at (2a): $(0,0,z)$ and $(0,0,z+1/2)$, Ba
positions are at (2b): $(1/3,2/3,z)$, $(2/3,1/3,z+1/2)$, and S
positions are at (6c): $(x,-x,z)$, $(x,2x,z)$, $(-2x,-x,z)$,
$(-x,x,z+1/2)$, $(-x,-2x,z+1/2)$, $(2x,x,z+1/2)$.  However, Niu's
positions \cite{niu2018} are not compatible with these positions.  For
instance, Niu's second set of the S position is at (0.8301, 0.6603,
0.850), but it should be (0.8301, 0.1699, 0.350). Their $y$ position
can be reproduced by $2x-1$, where $x$ is the $x$ position; also one
has to subtract 1/2 from their $z$ position.  In the following we
first correct their Wyckoff positions and then carry out the
calculation. Our Wyckoff positions are in compliance with the
international tables for crystallography.

We optimize both their structures, and find that the first structure
after optimization has a total energy lower than the second structure
by 4 mRy.  Both the structures have a lower energy than Singh's
structure. Our theoretical results support a structure with group
symmetry $\rm P6_3mc$.  This is the first testable case for future
experiments.  Our theoretically optimized Wyckoff positions, together
with the corrected experimental positions from Niu's paper
\cite{niu2018} are listed in Table \ref{our}.

\subsection{Electronic states}

Before we present the high harmonic generation spectrum of \Bae, we
first investigate its electronic structures.  Figure \ref{fig4}(a)
shows our total density of states (DOS). The Fermi energy ($E_f$) is
denoted by a dashed line. Consistent with Singh's results ($E_g=0.01$
eV) \cite{niu2018}, our energy gap is very small, around $E_g=0.014$
eV, on the energy scale of room temperature.  We expect some important
thermal-electric applications. Figure \ref{fig4}(b) shows the
element-resolved partial density of states. We notice that Ba has a
large contribution only in the lower energy window about 1 Ry below
the Fermi level.  Ti (dotted line) and S (dashed line) atoms dominate
DOS around the Fermi level. The partial DOS for S is shown in
Fig. \ref{fig4}(c), which is further resolved into different orbitals
in Fig. \ref{fig4}(d). It is clear that the states at -0.5 Ry are $3s$
states, while its $3p$ states are just around the Fermi level. As seen
below, these states provide a channel for HHG.

\subsection{High harmonic generation}

 We start with the structure with group symmetry $\rm P6_3/mmc$
 \cite{niu2018,huster1980}. We align the laser polarization along the
 $x$ axis.  Figure \ref{fig2}(a) shows the harmonic spectrum on a
 logarithmic scale as a function of harmonic orders. We see that all
 the harmonics appear at odd orders along the $x$ axis, which is the
 original laser polarization direction.  The signals along the $y$ and
 $z$ axes are at the noise level and not shown. Next, we use Niu's
 experimental structure with group symmetry $\rm P6_3mc$. Figure
 \ref{fig2}(b) shows that under the same laser condition, the harmonic
 signals along the $x$ axis for these two structures are identical
 (compare Figs. \ref{fig2}(a) and (b)), where all the harmonics are at
 odd orders. However, qualitative differences are observed along the
 $z$ axis. $\rm P6_3mc$ has no inversion symmetry and harmonics along
 the $z$ axis appear at even orders. This agrees with the symmetry
 properties with this group symmetry \cite{butcher,boyd} that the even
 orders only appear when the laser polarization is along the $z$ axis.
 Consistent with Niu's observation of strong optical anisotropy
 \cite{niu2018}, our zeroth-order harmonic signal along the $c$ axis
 is particularly strong; and to obtain clean harmonics at high orders,
 we subtract ${\bf P}(-\infty)$ from ${\bf P}(t)$ before we compute
 the power spectrum shown in Fig. \ref{fig2}(b). This is our second
 testable result: If Huster's structure is correct, no harmonic signal
 along the $z$ direction is present; if Niu's structure is correct,
 even harmonics along the $z$ axis appear.

\begin{figure}
  \includegraphics[angle=0,width=1\columnwidth]{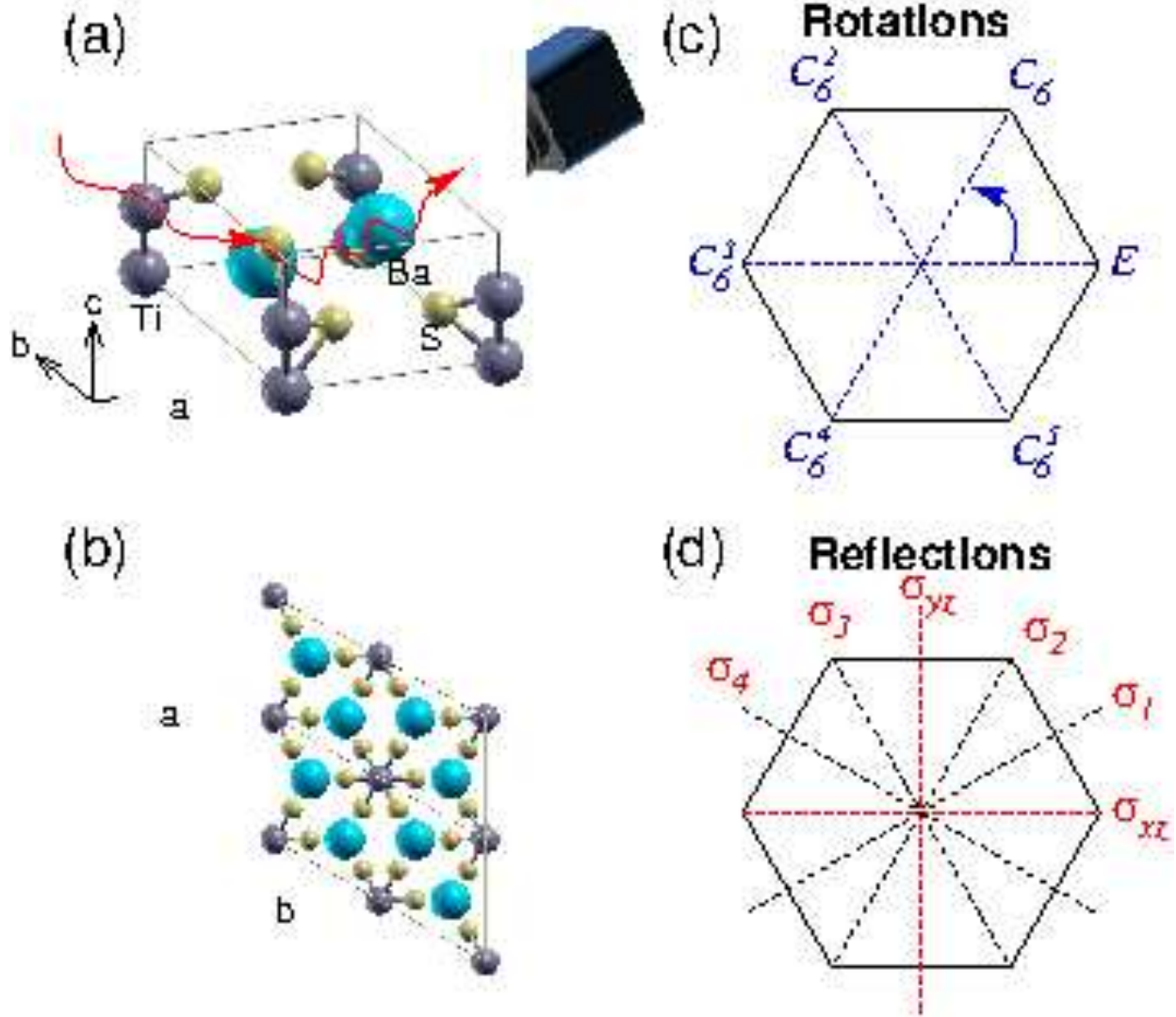}
  \caption
{High harmonic generation in \Bae. (a) Unit cell of \Ba with group
  symmetry $\rm P6_3mc$ has no inversion symmetry. The laser
  polarization can be either in the $xy$ $(ab)$ plane or along the $z$
  $(c)$ axis. Harmonics can be detected by a camera. 
(b) The same structure projected in the $ab$ plane, with
  two units along the $a$ and $b$ axes. (c) Six symmetry rotations in
  the $ab$ plane can be separated into two subgroups A ($E$, $C_6^3$)
  and B ($C_6$, $C_6^2$, $C_6^4$, $C_6^5$). These two subgroups are
  the cause of magic harmonics. 
(d) Six improper rotations
  can also be categorized into two subgroups A ($\sigma_{yz}$,
  $\sigma_{xz}$) and B ($\sigma_1$, ..., $\sigma_4$). }
\label{fig1}
  \end{figure}

\begin{figure}
  \includegraphics[angle=270,width=0.9\columnwidth]{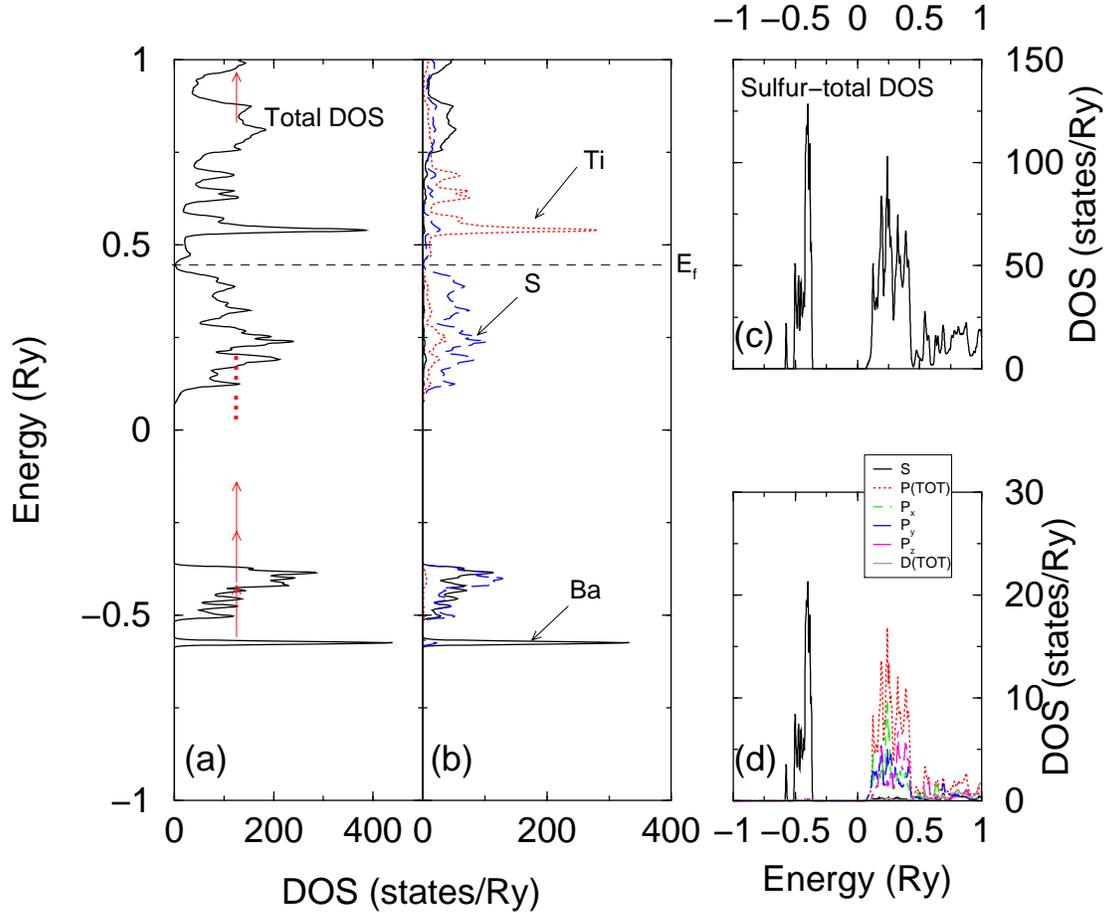}
  \caption
{(a) Total density of states (DOS). The arrows highlight the
  excitation process of our laser pulse. States from 41 to 146 are
  included in our calculation. The Fermi level is denoted by a
  horizontal dashed line.  (b) Element-resolved partial DOS, where the
  states around the Fermi level are dominated by Ti and S atoms.  (c)
  Partial density of states for sulfur, which is further decomposed
  into different orbital characters in (d).  }
\label{fig4}
  \end{figure}

\begin{figure}
  \includegraphics[angle=270,width=0.9\columnwidth]{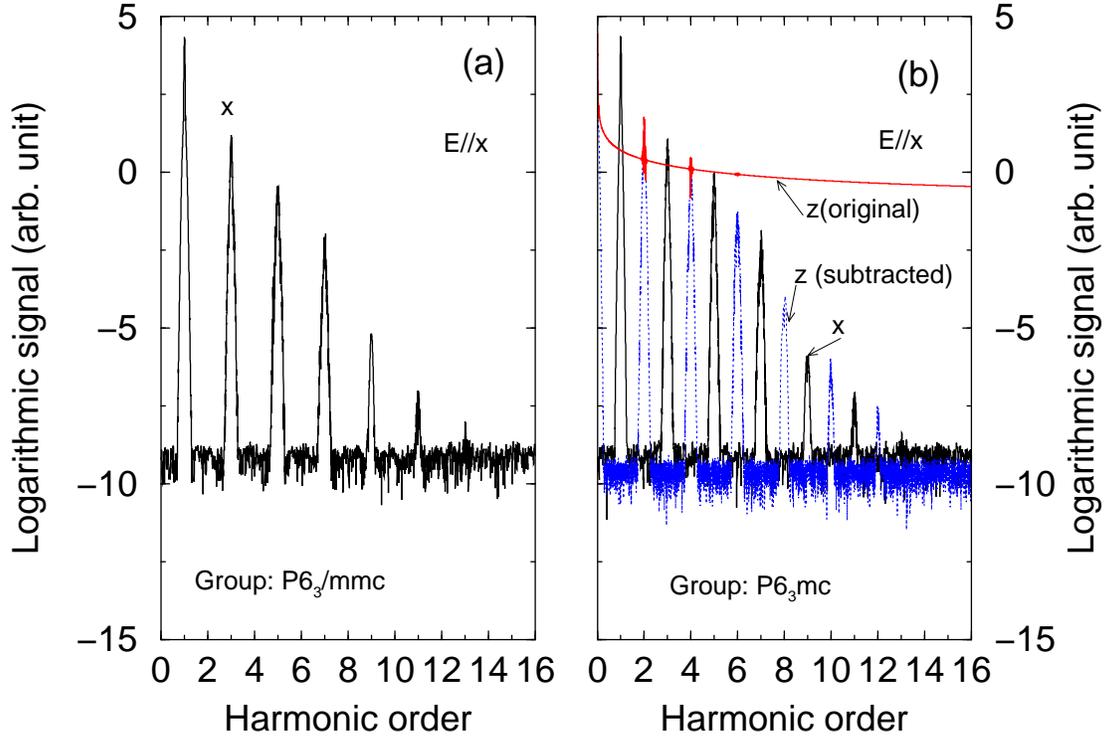}
  \caption
{(a) Logarithmic of high harmonic spectrum as a function of harmonic
  order under $x$-linearly polarized laser excitation. The crystal
  structure has group symmetry $\rm P6_3/mmc$ and has inversion
  symmetry. All the harmonics along the $x$ axis are odd
  orders. Signals along the other axes are at noise level.  (b)
  Harmonic signals for the same laser pulse but for the structure with
  group symmetry $\rm P6_3mc$. This structure has no inversion
  symmetry, so the even order harmonics appear along the $c$ ($z$)
  axis (dashed line). The signal along the $x$ axis is similar to (a).
  The $z$ component of harmonics without treatment is shown on the
  top. Since the signal along the $z$ axis is too strong, we subtract
  its initial value ${\bf P}(-\infty)$ from ${\bf P}(t)$ and then
  carry out the Fourier transformation to get a ``cleaner'' spectrum
  labeled with ``z (subtracted)''.  Even harmonic only appears when
  the laser polarization is along the $z$ axis \cite{butcher,boyd}. }
\label{fig2}
  \end{figure}

\begin{figure}
  \includegraphics[angle=270,width=0.9\columnwidth]{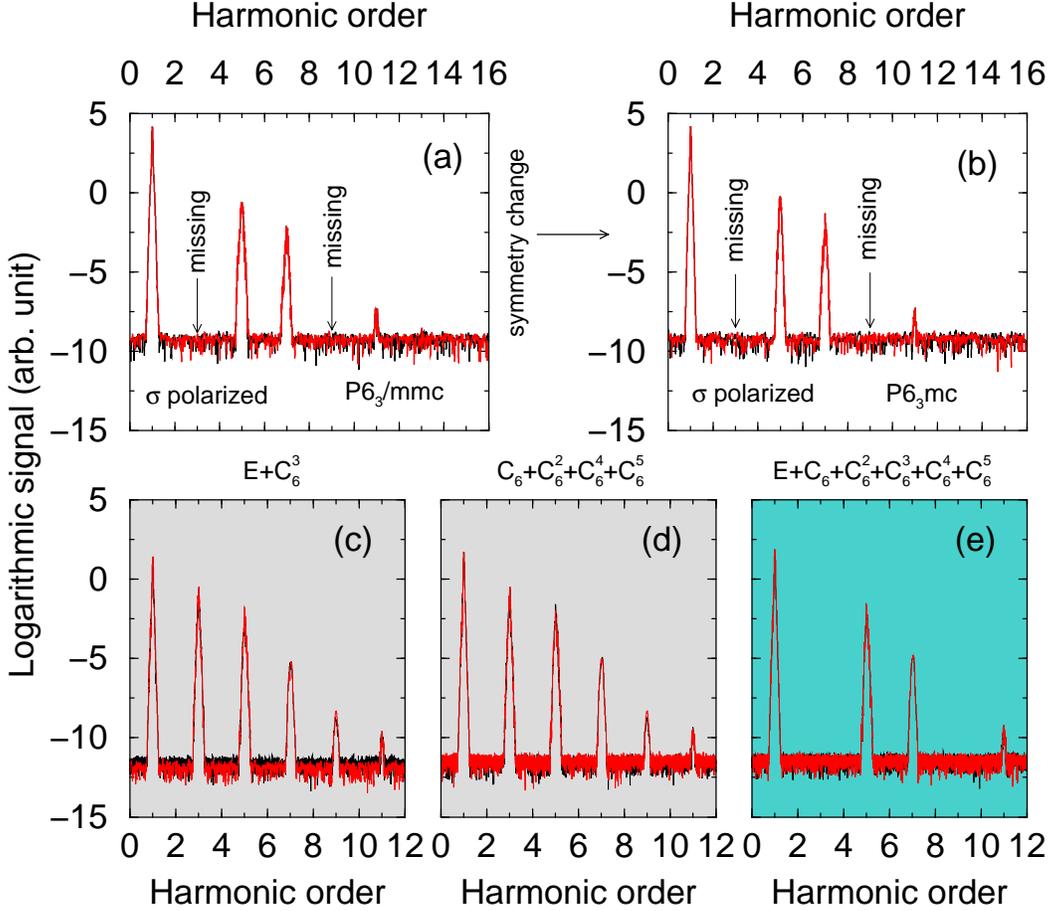}
  \caption
{(a) Magic high-order harmonics generated from \Ba with $\rm P6_3/mmc$
  symmetry under circularly polarized light ($\sigma$) in the $xy$
  plane, where third and ninth harmonics are missing.  The light
  polarization plane is within the $ab$ plane (see
  Fig. \ref{fig1}(b)).  Note that the $x$ and $y$ components are
  indistinguishable within numerical accuracy.  (b) Magic harmonics
  are also present in \Ba with a symmetry group $\rm P6_3mc$ without
  inversion symmetry. (c) High-order harmonics at a crystal momentum
  point close to (0.05, 0.05, 0).  The results are similar for other
  {\bf k} points. The summation is over symmetry operations in
  subgroup A ($E$ and $C_6^3$). The harmonics are normal.  (d) Same as
  (c), but the summation is over symmetry elements in subgroup B
  ($C_6$, $C_6^2$, $C_6^4$, $C_6^5$). These symmetry rotations keep Ti
  atoms intact while transforming S atoms into their equivalent
  positions (see Figs. \ref{fig1}(b) and \ref{fig1}(c)).  The
  harmonics are also normal and appear only at odd orders. (e)
  Summation of all the proper rotations on HHG signal reproduces magic
  harmonics seen in (a) and (b). It is the destructive interference
  between these two subgroups that cancels the third and ninth
  harmonics.  }
\label{fig3}
  \end{figure}

The structural symmetry is not the only information that HHG can
reveal. When we employ circularly polarized light ($\sigma$), to our
surprise, some harmonics are mysteriously missing. Figure
\ref{fig3}(a) shows that the third and ninth harmonics disappear. Only
the 1st, 5th, 7th, and 11th harmonics remain, magic harmonics.  To the
best of our knowledge, this has never been reported before. These
magic harmonics do not depend on whether the group symmetry is $\rm
P6_3/mmc$ (Fig. \ref{fig3}(a)) or $\rm P6_3mc$
(Fig. \ref{fig3}(b)). Therefore, the common symmetry operations shared
by these two space groups must be at the root of these magic
harmonics. However, since we have a huge number of {\bf k} points, it
is a challenge to determine the origin of these magic harmonics. We
decide to select a single {\bf k} point and resolve ${\bf P}_{\bf
  k}(t)$ according to its 12 symmetry operations \be {\bf P}_{\bf
  k}(t)=\sum_{s=1}^6{\bf P}_{\bf k}^s(t)+\sum_{q=1}^6{\bf P}_{\bf
  k}^q(t) \label{eq3} \ee where $s$ refers to six proper rotations
(see Fig. \ref{fig1}(c)) and $q$ runs over six improper rotations
(reflections) (Fig. \ref{fig1}(d)). We symmetry-resolve ${\bf P}_{\bf
  k}(t)$, not its Fourier transformed ${\bf P}_{\bf k}(\omega)$,
because the interference only occurs in the time domain, not in the
frequency domain.  Equation (\ref{eq3}) is seemingly simple, but
harbors too many possible combinations, $ \sum_{i=1}^{12} \left
( \begin{array}{c} i\\ 12\\
\end{array}
\right ).
$
Lax \cite{lax} has an example of an equilateral triangle, with six
symmetry operations. His example can not directly apply to our
problem, but we notice that in his diagram the triangle has a mirror
plane which could cancel all the even-order harmonics. This motivates
us to lay out all the proper rotations within the $ab$ plane
(Fig. \ref{fig1}(c)), where we label each vertex with a symmetry
operation.  $C_6^3$ is a $180^\circ$ rotation with respect to the $c$
axis, much like an inversion operation in the Lax's example.  We
immediately recognize that the identity matrix $E$ and rotation
$C_6^3$, or subgroup A below, ensure that even harmonics do not
appear.  This is verified by our calculation (see Fig. \ref{fig3}(c)),
but their harmonics are normal, and no magic orders are observed. Note
that the disappearance of the even harmonics does not contradict the
symmetry properties because the even harmonics allowed by the symmetry
$\rm P6_3 mc$ appears only when the electric field is along the $z$
axis \cite{butcher,boyd}. In our case, our laser field polarization is
in the $xy$ plane. Experimentally, Ghimire \et \cite{ghimire2011}, who
employed linearly polarized in the $ab$ plane, also found no even
order harmonics for ZnO which happens to have the same group symmetry
$\rm P6_3 mc$. Therefore, our results are fully consistent with the
symmetry requirement and prior experiments \cite{ghimire2011}.

What about the remaining four proper rotations or subgroup B? If we
compare Figs. \ref{fig1}(b) and \ref{fig1}(c), we notice that these
symmetry operations bring S atoms to their equivalent positions while
keeping Ti atoms intact. These four rotation matrices are
\begin{equation}
C_6(60^\circ)=
\left(
\begin{array}{rrr}
\frac{1}{2} & \frac{\sqrt{3}}{2}  &~~ 0\\
 -\frac{\sqrt{3}}{2} &\frac{1}{2} & 0\\
 0 & 0 & 1\\
\end{array}
\right ), \hspace{2cm}
C_6^2(120^\circ)=\left(
\begin{array}{rrr}
-\frac{1}{2} & \frac{\sqrt{3}}{2}  &~~ 0\\
 -\frac{\sqrt{3}}{2} &-\frac{1}{2} & 0\\
 0 & 0 & 1\\
\end{array}
\right ),
\label{sym1a}
\label{sym1b}
\end{equation}
\begin{equation}
C_6^4(240^\circ)=
\left(
\begin{array}{rrr}
-\frac{1}{2} & -\frac{\sqrt{3}}{2}  &~~ 0\\
 \frac{\sqrt{3}}{2} &-\frac{1}{2} & 0\\
 0 & 0 & 1\\
\end{array}
\right ), 
\hspace{2cm}
C_6^5(300^\circ)=\left(
\begin{array}{rrr}
\frac{1}{2} & -\frac{\sqrt{3}}{2}  &~~ 0\\
 \frac{\sqrt{3}}{2} &\frac{1}{2} & 0\\
 0 & 0 & 1\\
\end{array}
\right ).
\label{sym2}
\end{equation}
\noindent
We first check whether each pair of symmetry operations lead to magic
harmonics, but this fails, so we simply combines all of them. The
results are shown Fig. \ref{fig3}(d), where all the harmonics appear
normal. Then we test whether mixing some improper rotations could
cause magic harmonics, but unsuccessful. We also examine the prior
results in magnetic monolayers with tetragonal symmetries \cite{nc18},
but we do not find any magic harmonics under the same laser
condition. After a long and difficult testing, we finally come to
realize that the summation over all ${\bf P}_s(t)$ of the proper
rotation might be able to reveal magic harmonics, and to our
amazement, it indeed works.  Figure \ref{fig3}(e) shows magic
harmonics, with the third and ninth order harmonics missing. It is the
destructive interference between two symmetry subgroups A and B that
leads to magic harmonics. It is an easy task to extend this finding to
improper rotations (see Fig. \ref{fig1}(d)). Different from the proper
rotations, the mirror image operations $\sigma_{yz}$ and $\sigma_{xz}$
form a subgroup, with the four remaining operations from $\sigma_1$ to
$\sigma_4$ forming another subgroup. These two subgroups play the same
role as the two subgroups for the proper rotations. Once we add their
contributions up, we again reproduce the magic harmonics. We also
examine other hexagonal systems and reach the same conclusion. These
magic harmonics are a hallmark of hexagonal structure, which is likely
to have important applications in the future.  For instance,
hexagonal-cubic crystal structure transformation was found in many
technologically important materials, aluminum nitride
\cite{iwata2005}, GaN \cite{menniger1996}, BN
\cite{milledge1959,olander2003,halo2011}, zinc oxynitride layers
\cite{gomez-castano2018}, NaYF$_4$ \cite{saha2018}, Eu$_2$O$_3$
\cite{concas2011}, CdTe \cite{ariz1997} and others
\cite{castedo2011}. Our finding suggests a simple protocol to
determine whether a hexagonal-cubic phase transition occurs by
checking whether these magic harmonics appear.

\section{Conclusion}

We have demonstrated magic high-order harmonics in hexagonal and
quasi-one dimensional solid \Bae. Our results reflect the usefulness
of group theory and power of high harmonic generation as a structural
characterization tool. Specifically, we show how harmonics are
generated sensitively depends on crystal structures and laser
polarization. Whether \Ba adopts $\rm P6_3/mmc$ or $\rm P6_3mc$
symmetry determines whether even order harmonics appear along the $c$
axis.  The qualitative difference is found under circularly ($\sigma$)
polarized light excitation between the hexagonal structure for \Ba and
tetragonal structure \cite{nc18}. $\sigma$ light produces no magic
harmonics in tetragonal systems, but it generates magic harmonics in
hexagonal systems.  These magic harmonics are the hallmark of the
hexagonal structure, and potentially provide a tool to investigate
phase transitions in a wide scope of materials. One ideal system to
realize our prediction could be BaVS$_3$. At room temperature BaVS$_3$
adopts $\rm P6_3/mmc$ symmetry, but transforms to an orthorhombic
structure between 70 K and 240 K \cite{fagot2005}. Therefore, our
finding will motivate experimental and theoretical investigations in
other research fields.

\acknowledgments We would like to thank Dr. D. Singh (University of
Missouri) for sending us his structure file.  This work was solely
supported by the U.S. Department of Energy under Contract
No. DE-FG02-06ER46304. Part of the work was done on Indiana State
University's high performance Quantum and Obsidian clusters.  The
research used resources of the National Energy Research Scientific
Computing Center, which is supported by the Office of Science of the
U.S. Department of Energy under Contract No. DE-AC02-05CH11231.


\begin{table}
\caption{Wyckoff positions of BaTiS$_3$ determined by Huster
  \cite{huster1980}.  His structure is of BaNiO$3$-type and has
  symmetry group No. 194, $\rm P6_3/mmc$.  This structure has an inversion
  symmetry.  The lattice constants are $a=6.756(1)\rm\AA$ and
  $c=5.798(1)\rm \AA$.  }
\begin{tabular}{cclll}
\hline\hline
Atom & Position & $x$ & $y$ &$z$ \\
\hline
Ti &2a & 0& 0& 0 \\
Ba& 2d & $\frac{1}{3}$ &$\frac{2}{3}$ & $\frac{3}{4}$\\
S &6h & 0.1655(10) & 0.3310(10) & 1/4\\
\hline\hline
\end{tabular}
\label{huster}
\end{table}

\begin{table}
\caption{Optimized Wyckoff positions of BaTiS$_3$, with group No. 186,
  $\rm P6_3mc$ and lattice constants $a=6.749\rm \AA$, $b=6.749\rm
  \AA$, and $c=5.831\rm \AA$.  S atoms are at $6c$ positions $(x,-x,z)$,
  $(x,2x,z)$, $(-2x,-x,z)$, $(-x,x,z+\frac{1}{2})$,
  $(-x,-2x,z+\frac{1}{2})$, and $(2x,x,z+\frac{1}{2})$.  This
  structure has no inversion symmetry.  The experimental results from
  Niu \et \cite{niu2018} are shown in the parenthesis. If there is no
  difference between their experiment and our theory, only one entry
  is listed.  }
\begin{tabular}{cclll}
\hline\hline
Atom & Position & $x$ & $y$ &$z$ \\
\hline
Ba& 2b & $\frac{1}{3}$ &$\frac{2}{3}$ & 0.2989 (0.298)\\
Ti &2a & 0& 0& 0.5067 (0.522) \\
S &6c & 0.8329 (0.8301) & 0.1671 (0.1618) & 0.2875 (0.298)\\
\hline\hline
\end{tabular}
\label{our}
\end{table}

\end{document}